\documentclass[letterpaper, 10 pt, conference]{ieeeconf}  

\IEEEoverridecommandlockouts                              

\usepackage{amsmath,amssymb,bm} 
\usepackage{acronym}
\usepackage{color}
\usepackage{hyperref}
\usepackage{graphicx}		
\usepackage{mathtools}
\usepackage[normalem]{ulem}
\usepackage{comment}
\usepackage{algorithm,algpseudocode}
\usepackage{cite}
\usepackage{xcolor}
\usepackage{cleveref}
\usepackage{subcaption}
\usepackage{tikz}

\definecolor{mygray}{gray}{0.35}

\newtheorem{definition}{Definition}[section]

\newcommand{\defeq}{\vcentcolon=}

\newtheorem{theorem}{Theorem}

\newtheorem{lemma}{Lemma}

\newtheorem{assumption}{Assumption}

\newcommand{\Scal}{\mathcal{S}} 
\newcommand{\mbf}[1]{\mathbf{#1}} 
\newcommand{\bmat}[1]{\begin{bmatrix} #1 \end{bmatrix}} 
\newcommand{\Xcal}{\mathcal{X}}

\newcommand{\xhat}{\hat{\mathbf{x}}} 
\newcommand{\x}{\mathbf{x}} 
\newcommand{\bu}{\mathbf{u}} 
\newcommand{\y}{\mathbf{y}} 
\newcommand{\z}{\mathbf{z}} 
\newcommand{\Q}{\mathbf{Q}} 

\newcommand{\X}{\mbf{X}}

\newcommand{\vecdim}[1]{\in \mathbb{R}^{#1}}

\newcommand{\norm}[1]{\left\lVert#1\right\rVert}
\newcommand{\absVal}[1]{\left\lvert#1\right\rvert}

\newcommand{\Pre}{\text{Pre}}
\newcommand{\Suc}{\text{Suc}}

\newcommand{\Qcal}{\mathcal{Q}}

\newcommand{\Dcal}{\mathcal{D}}

\DeclareMathOperator*{\argmin}{arg\,min}

\Crefname{algorithm}{Alg.}{Algs.}

\acrodef{dl}[{DL}]{deep learning}
\acrodef{rl}[{RL}]{reinforcement learning}
\acrodef{nn}[{NN}]{neural network}
\acrodef{dnn}[{DNN}]{deep neural network}
\acrodef{tdl}[{TDL}]{temporal difference learning}
\acrodef{pid}[{PID}]{proportional–integral–derivative}
\acrodef{us}[{US}]{Ultrasound}
\acrodef{mse}[{MSE}]{mean squared error}
\acrodef{sgd}[{SGD}]{stochastic gradient descent}
\acrodef{ico}[{ICO}]{iterative convex overbounding}
\acrodef{lmi}[{LMI}]{linear matrix inequality}
\acrodef{mjls}[{MJLS}]{Markov jump linear system}
\acrodef{io}[{IO}]{input-output}
\acrodef{iqc}[{IQC}]{integral quadratic constraints}
\acrodef{cnn}[{CNN}]{convolutional neural network}
\acrodef{il}[{IL}]{Imitation learning}
\acrodef{mpc}[{MPC}]{model predictive control}
\acrodef{sdp}[{SDP}]{semi-definite programming}
\acrodef{relu}[{ReLU}]{rectified linear unit}
\acrodef{us}[US]{ultrasound}
\acrodef{mdp}[MDP]{Markov Decision Process}
\acrodef{iid}[iid]{identical and independently distributed random variable}
\acrodef{pid}[PID]{Proportional Integral Derivative}

\acrodef{lqr}[LQR]{linear-quadratic regulator}

\acrodef{cpa}[CPA]{continuous piecewise affine}
\acrodef{hji}[HJI]{Hamilton Jacobi Inequality}
\acrodef{pi}[PI]{positive invariant}
\acrodef{gaio}[GAIO]{Global Analysis of Invariant Objects}

\title{\LARGE \bf
Synthesizing provably invariant sets via stochastically sampled data
}

\author{Amy K. Strong$^{1}$, Ali Kashani$^{2}$, Claus Danielson$^{3}$, and Leila Bridgeman$^{1}$
\thanks{$^{1}$Amy K. Strong and Leila J. Bridgeman are with the Department of Mechanical Engineering and Materials Science at Duke University, Durham, NC, 27708, USA.
        (email: {\tt\small aks121@duke.edu,  ljb48@duke.edu})}%
\thanks{$^{2}$Ali Kashani is with the School of Data Science at the University of Virginia, Charlottesville, VA, 22903, USA.
        (email: {\tt\small fww9ba@virginia.edu})}
\thanks{$^{3}$Claus Danielson is with the Department of Mechanical Engineering at the University of New Mexico, Albuquerque, NM, 87131, USA.
        (email: {\tt\small cdanielson@unm.edu})}}%

\newcommand\copyrighttext{%
  \footnotesize \textcopyright 2026 IEEE. Personal use of this material is permitted.
  Permission from IEEE must be obtained for all other uses, in any current or future
  media, including reprinting/republishing this material for advertising or promotional
  purposes, creating new collective works, for resale or redistribution to servers or
  lists, or reuse of any copyrighted component of this work in other works.}
  \newcommand\copyrightnotice{%
\begin{tikzpicture}[remember picture,overlay]
\node[anchor=south,yshift=10pt] at (current page.south) {\fbox{\parbox{\dimexpr\textwidth-\fboxsep-\fboxrule\relax}{\copyrighttext}}};
\end{tikzpicture}%
}

\begin{document}

\maketitle
\copyrightnotice
\thispagestyle{empty}
\pagestyle{empty}

\begin{abstract}

\Ac{pi} sets are essential for ensuring safety, i.e. constraint adherence, of dynamical systems. With the increasing availability of sampled data from complex (and often unmodeled) systems, it is advantageous to leverage these data sets for \ac{pi} set synthesis. This paper uses data driven geometric conditions of invariance to synthesize \ac{pi} sets from data. Where previous data driven, set-based approaches to \ac{pi} set synthesis used deterministic sampling schemes, this work instead synthesizes \ac{pi} sets from any pre-collected data sets. Beyond a data set and Lipschitz continuity, no additional information about the system is needed. A tree data structure is used to partition the space and select samples used to construct the \ac{pi} set, while Lipschitz continuity is used to provide deterministic guarantees of invariance. Finally, probabilistic bounds are given on the number of samples needed for the algorithm to determine of a certain volume.

\end{abstract}

\section{Introduction}

A \acf{pi} set is a subset of the state space where, once entered, a dynamical system remains for all time. Determining a \ac{pi} set within a state constraint set results in a region where the system adheres to constraints in perpetuity. As such, a large body of work on the safety of dynamical systems is dedicated to determining \ac{pi} sets. The two most influential model-based approaches are the seminal geometric algorithm \cite{kerrigan2000robust, gilbert1991linear}, which determines the \ac{pi} set of a system through an iterative set based approach, and barrier function based methods \cite{lasalle2012stability,alberto2007invariance}, which determine a system's scalar energy function -- the sublevel set of which denotes a \ac{pi} set. The geometric algorithm is particularly influential as it, under certain assumptions, guarantees finding the maximal \ac{pi} set of a system \cite{gilbert1991linear} in finite time. 

Increasingly, vast amounts of data are collected from complex -- and potentially unmodeled -- systems. As a result, there is focus on determining \ac{pi} sets via data driven methods. When the unmodeled system is assumed to be linear, William's fundamental lemma \cite{willems2005note} is often used to determine the \ac{pi} set of a system \cite{mulagaleti2021data,mejari2023direct}. For unmodeled systems that have nonlinear dynamics, current literature primarily focuses on synthesizing barrier functions for the system \cite{korda2020computing,anand2023formally,zhang2024seev,edwards2024fossil,richards2018lyapunov,zhang2023efficient,robey2020learning}. For these methods, the model is either known \cite{zhang2023efficient,edwards2024fossil,richards2018lyapunov,robey2020learning} or approximated via system identification techniques and the barrier function is synthesized using the learned model \cite{jagtap2020control,richards2018lyapunov}. Alternately, the barrier function can be learned directly from data without system identification \cite{jin2023robust,korda2020computing,anand2023formally}. 



To the best of our knowledge, there has been little exploration into data driven geometric methods of determining \ac{pi} sets. Of the work available, probabilistic -- rather than deterministic -- guarantees of positive invariance are established \cite{wang2020scenario} or the method requires a deterministic sampling scheme \cite{strong2025data,strong2025certificates}. For safety critical systems, probabilistic guarantees of invariance (i.e., the probability of a state in the \ac{pi} set violating the invariance condition is less than $\epsilon$ \cite{wang2020scenario}) may not be enough. Instead, it may be desirable to have deterministic guarantees (i.e., all states in the \ac{pi} set will always adhere to the invariance condition). However, the expectation of deterministic sampling capabilities is not always realistic. The aim of this paper, then, is to leverage pre-existing data sets collected from an unmodeled system to determine a provably \ac{pi} set using a data-driven, geometric method. This requires an additional assumption that the system is Lipschitz continuous with a known Lipschitz constant, a common assumptions in the literature \cite{anand2023formally,robey2020learning,richards2018lyapunov}.

We build upon previous work \cite{strong2025certificates}, which was inspired by the seminal geometric method \cite{kerrigan2000robust,gilbert1991linear} and the subdivision algorithm \ac{gaio} \cite{dellnitz2001algorithms}. In \cite{strong2025certificates}, the state constraint set was partitioned via max-norm balls centered at deterministically sampled states, $\x$. The partitions were iteratively labeled based on an over-approximation of their behavior one-step forward in time. Now, we instead over-approximate the one-step forward behavior of a partition via nearby samples in the pre-collected data set. The result is an algorithm capable of producing a \ac{pi} set which approximates the maximal \ac{pi} set using any available data set. By leveraging Lipschitz continuity, deterministic guarantees of positive invariance are ensured. Our algorithm requires no additional knowledge about the system beyond Lipschitz continuity and a sampled data set of state and successor pairs, $\Dcal = \{\x_j,\x_j^+\}_{j=1}^M.$ 

While our algorithm is guaranteed to produce a \ac{pi} set, there may be cases, such as sparse data sets or systems with a high valued Lipschitz constant, where the \ac{pi} set produced is the null set. Therefore, we also provide analysis on the sampling bounds for our algorithm to produce a non-empty \ac{pi} set. Specifically, we leverage tools from set coverage analysis to determine the probability of our algorithm producing a non-empty \ac{pi} set based on the samples provided.

\subsection*{Notation and Preliminaries} 
Let $\mathbb{Z}_a^b$ be the set of integers between $a$ and $b$ inclusive. 
Scalars, vectors, and matrices are denoted as $x,$ $\x,$ and $\X$. Let $\{1, -1\}^n$ denote the set of all $n$-dimensional vectors with entries either $1$ or $-1$.

Define a norm ball a $\x \vecdim{n}$ for some norm, $\norm{\cdot}_p$ and $p \in \mathbb{Z}_1^\infty$, as $B_{r,p}(\x){\defeq} \{\y \vecdim{n} {\mid} \norm{\x-\y}_p{\leq} r\}$ or, equivalently, $B_{r,p}(\x) \defeq \{\x + r\bu \mid \norm{\bu}_p \leq 1\}$ \cite{boyd2004convex}. If $p$ is not specified, then any norm $p \in \mathbb{Z}_1^\infty$ applicable.  Recall that the max norm ($p=\infty$) is defined as $\norm{\x}_\infty {= }\max_{i \in \mathbb{Z}_1^n} \absVal{x_i}$. Define an $\epsilon$-net of a set $\Theta \subseteq \mathbb{R}^n$ with respect to a norm, $\norm{\cdot}_p$, as a set $\{\x_i\}_{i=1}^N\in \Theta$ such that for every $\y\in\Theta$, there exists some $i \in \mathbb{Z}_1^N$ such that $\norm{\x_i-\y}_p \leq \epsilon$ \cite{wainwright2019high}.

Let a C-set be defined as a convex and compact subset of $\mathbb{R}^n,$ which includes the origin as an interior point. Recall that, given a C-set $\Scal$, a Minkowski function \cite[Def. 3.11]{blanchini2015set} is defined as $\psi_\Scal(\x) = \inf\{\lambda \geq 0 \mid \x \in \lambda \Scal\}.$ The Minkowksi function is convex, positively homogenous of order one, and sub-additive \cite[Prop. 3.12]{blanchini2015set}. Let $\oplus$ denote the Minkowski sum, defined as $\mathcal{A} \oplus \mathcal{B} \defeq \{a+b \mid a\in\mathcal{A}, b\in\mathcal{B}\}$ \cite[Def. 3.8]{blanchini2015set}. Finally, let $\text{vol}(.)$ return the volume of set.

The mapping $T{:}\Xcal {\rightarrow} \mathcal{Y}$ between two metric spaces is Lipschitz continuous with respect to the norm $\norm{\cdot}_p$ if there exists some Lipschitz constant, $L {>} 0$, such that $\norm{T(\mathbf{p}) {-} T(\mathbf{q})}_p {\leq} L \norm{\mathbf{p} {-} \mathbf{q}}$ for all $\mathbf{p}, \mathbf{q} \in \Xcal$\cite{fitzpatrick2009advanced}. Let $T^k$ indicate the mapping is applied $k$ times ($k{\in}\mathbb{Z}_{0}^{\infty}$).

\section{Problem Statement}

Consider a (potentially unmodeled) discrete-time, dynamical system
\begin{flalign}\label{eq:dynSys}
    \x^+ = T(\x)
\end{flalign}
over the bounded state constraint admissible set, $\Xcal \subset \mathbb{R}^n.$ The following assumption is made.
\begin{assumption}\label{asmpt:Lipschitz}
    Let \eqref{eq:dynSys} be Lipschitz continuous for some norm, $\norm{\cdot}_p.$ Let $L>0$ be an upper bound on the Lipschitz constant of the system.
\end{assumption}

Consider a data set $\Dcal = \{\x_j,\x_j^+\}_{j=1}^M$ of $M$ states $\x_i\in\Xcal$ and their corresponding successor state, $\x_i^+ = T(\x_i).$ This work aims to use this pre-collected data set to determine a \ac{pi} set which can approximate the maximal \ac{pi} set within $\Xcal$.


\subsection{Geometric conditions of positive invariance}

A \ac{pi} set is a region of the state space that the system will remain within indefinitely, as defined below.
\begin{definition}[\ac{pi} Set\cite{alberto2007invariance}]\label{def:invSet}
    $\Scal$ is \ac{pi} under the dynamics of \eqref{eq:dynSys} if, $ \forall\x_0 \in \Scal, k \in\mathbb{Z}_{0}^{\infty}$, $T^k(\x_0) \in \Scal.$
\end{definition}
A stricter condition than positive invariance is a set being $\lambda$-contractive. This requires that, once a system enters a region, it maps to a smaller subset of that region. Contractivity can be defined with respect to the Minkowski function.
\begin{definition}[$\lambda-$contractive Set \cite{blanchini2015set}]\label{def:contSet}
    Let $0\leq \lambda < 1.$ The C-set $\Scal$ is $\lambda-$contractive under the dynamics of \eqref{eq:dynSys} if and only if $\psi_\Scal(\x^+) \leq \lambda$ for every $\x \in\Scal$.
\end{definition}

Positive invariance can be understood geometrically. The precursor and successor set define subsets of the state space with respect to the evolution of a system one step in time.
\begin{definition}[Precursor Set \cite{alberto2007invariance}]
\label{def:precursor}
	For $T: \mathbb{R}^n \rightarrow \mathbb{R}^n,$ the precursor set to set $\Scal$ is $\Pre(\Scal) =\{\x\vecdim{n}\mid T(\x) \in \Scal\}.$
\end{definition}
\begin{definition} [Successor Set \cite{borrelli2017predictive}]
\label{def:successor}
	For $T: \mathbb{R}^n \rightarrow \mathbb{R}^n,$ the successor set to set $\Scal$ is $\Suc(\Scal) = \{\x \vecdim{n} \mid \exists \x_0\in \Scal \text{ s.t. } \x = T(\x_0)\}.$ 
\end{definition}
Then, a \ac{pi} set can be defined as a subset of the state space that maps into itself one step in time.
\begin{lemma}[\ac{pi} Set\cite{alberto2007invariance,dorea1999b}]\label{lem:invSet}
    The set $\Scal \subseteq \Omega$ is \ac{pi} for mapping $T: \mathbb{R}^n \rightarrow \mathbb{R}^n,$ if and only if $\Suc(\Scal)\subseteq \Scal$ ($\Scal \subseteq \Pre(\Scal)$).
\end{lemma}

\Cref{lem:invSet} is used in the seminal, model-based geometric algorithm that determines a system's maximal \ac{pi} set in a given state constraint set \cite{kerrigan2000robust,gilbert1991linear}.
In it, the entire constraint set is first assumed to be a \ac{pi} set ($\Scal \leftarrow \Xcal$). Then, the candidate \ac{pi} set is iteratively pruned using the one-step evolution of the system ($\Scal \leftarrow \Pre(\Scal)\cap \Scal$) until it remains unchanged from iteration to iteration, i.e. $\Scal = \Pre(\Scal) \cap \Scal$. 

While this algorithm is a cornerstone method for safety in dynamical systems, it can be challenging to implement depending on the circumstances.
Determining the precursor set requires a model, which may be unavailable. Even with a model, it can be difficult to compute the precursor set with a nonlinear model or a nonconvex state constraint set. 

\subsection{Data driven geometric conditions of invariance}\label{sec:geo}

Rather than directly computing the precursor and successor sets, \cite{strong2025data} proposed using Lipschitz continuity of the system to over-approximate these sets about individual sampled states -- circumventing the need for a system model. First, a norm ball describes the region about a sampled point. Then, Lipschitz continuity is used to understand the system evolution of that region over time, as seen in \Cref{lem:sucPrec}.

\begin{lemma}[\cite{strong2025data}]\label{lem:sucPrec}
    Let \Cref{asmpt:Lipschitz} hold. Consider the point $\x {\in} \Xcal$ and its successor, $\x^+ {=}T(\x).$  Define $B^+_r(\x) {\defeq }\{\y\vecdim{n} \mid \norm{\x^+ {-} \y}{\leq} L r\}$. Then, $ \Suc(B_r(\x)) {\subseteq} B^+_r(\x),$ and $B_r(\x) {\subseteq} \Pre(B^+_r(\x)).$
\end{lemma}
As a result of \Cref{lem:invSet}, a \ac{pi} set can then be defined using sets sampled data pairs, $\{\x,\x^+\}_{i=1}^N.$
\begin{lemma}[\cite{strong2025data}]\label{lem:pointsInv}
    Let \Cref{asmpt:Lipschitz} hold. Consider a data set of $N$ pairs, $\{\x_i, \x^+_i\}_{i=1}^N$, sampled in $\Xcal {\subset} \mathbb{R}^n$ where each element in a pair are related via \eqref{eq:dynSys}. Let $B_{r_i}(\x)$ and $B^+_{r_i}(\x_i)$ be as defined in \Cref{lem:sucPrec}.
    If $\cup_{i=1}^N B^+_{r_i}(\x_i) \subseteq \cup_{i=1}^NB_{r_i}(\x_i),$ then  $\cup_{i=1}^NB_{r_i}(\x_i)$ is a \ac{pi} set. 
\end{lemma}

\Cref{lem:sucPrec} and \Cref{lem:pointsInv} have been leveraged to create data driven geometric algorithms to determine a system's \ac{pi} set \cite{strong2025data,strong2025certificates}. However, these algorithms require deterministically sampling data pairs $\{\x,\x^+\}$. Requiring repeated (and noiseless) interactions with a dynamical system is often not realistic. Further, many systems may already have pre-collected data sets that have been used for other analysis methods. It is therefore advantageous to adapt previous methods \cite{strong2025certificates} to be able to handle any data set available.

\section{Algorithm for Data Driven PI Set Synthesis}


We leverage data driven geometric conditions for \ac{pi} sets (\Cref{lem:pointsInv}) to develop \Cref{alg:findSet}, which determines an invariant approximation of the maximal \ac{pi} set of a dynamical system \eqref{eq:dynSys} given a state constraint set, $\Xcal$, any sampled data set, $\Dcal = \{\x_j,\x_j^+\}_{j=1}^M,$ and an overbound on the Lipschitz constant of the system, $L.$ Unlike deterministic sampling algorithms  \cite[Alg. 1]{strong2025certificates},  \Cref{alg:findSet} can handle pre-collected data sets.



Broadly, \Cref{alg:findSet} uses a tree data structure to partition $\Xcal$ using hypercubes and label each partition as part of (or not part of) a candidate \ac{pi} set. 
To do so, each partition is matched to a sampled pair $\{\x,\x^+\}$ in $\Dcal$ and labeled using the behavior of the sampled data pair.
\Cref{sec:part} details the tree data structure and its relationship to the partitions, while \Cref{sec:algTotal} describes the overall algorithm for finding provably \ac{pi} sets.

\subsection{Tree data structure}\label{sec:part}

\Cref{def:Q} describes the tree structured used to store and label partitions of $\Xcal$.
%
%

\begin{definition} \label{def:Q}
    Define $\Qcal = \Q(\hat{r}_i, \hat{\x}_i, r_i, \{\x_i, \x_i^+\}, s_i)_{i=1}^N$ as a tree data structure containing $N$ nodes, $\Q(\hat{r}_i, \hat{\x}_i, r_i, \{\x_i, \x_i^+\}, s_i)$.
    Each node contains the target radius $\hat{r}_i$ and target center, $\hat{\x}_i$ that create the hypercube partition $B_{\hat{r}_i,\infty}(\xhat_i){\subset} \Xcal.$ 
    The sampled data pair $\{\x_i,\x^+_i\}$ is a sampled state and successor from the data set, $\Dcal$, where the value $r_i$ describes radius of the tightest max-norm ball, $B_{r_i,\infty}(\x_i)\supseteq B_{\hat{r}_i,\infty}(\xhat_i).$
    Label $s_i$ denotes if partition $B_{\hat{r}_i,\infty}(\xhat_i)$ is included in ($s_i{=}1$) or excluded from ($s_i{=}0$) a candidate \ac{pi} set, $\hat{\Scal}$, or if its status is unknown ($s_i{=}-1$).  
\end{definition}

%


\Cref{alg:divide} describes dividing a node of $\Qcal$ and ensuring that $B_{r_i,\infty}(\x_i){\supseteq} B_{\hat{r}_i,\infty}(\xhat_i)$ for each new node, as stated in \Cref{lem:overPart}.

\begin{lemma}\label{lem:overPart}
    Let $\Qcal$, node $i$, and the data set $\Dcal$ be inputs to \Cref{alg:divide}. Then, $B_{r_k,\infty}(\x_k){\supseteq} B_{\hat{r}_k,\infty}(\xhat_k)$ for each $k{\in} \mathbb{Z}_{N+1}^{N+1+2^n}.$
\end{lemma}
\begin{proof}
    Any point $z \in B_{\hat{r}_k,\infty}(\xhat_k)$ can be defined as $\xhat_k +\hat{r}_k\bu$, where $\norm{\bu}_\infty \leq 1.$ Consider then $\norm{\x_k - \xhat_k +\hat{r}_k\bu}_\infty$.  By the triangle inequality, this is bounded above by $\norm{\x_k -\xhat_k}_\infty + \hat{r}_k\norm{\bu}_\infty \leq \norm{\x_k -\xhat_k}_\infty + \hat{r}_k \leq r_k.$ Thus, $B_{r_k,\infty}(\x_k)\supseteq B_{\hat{r}_k,\infty}(\xhat_k)$.
\end{proof}

\begin{algorithm}
    \caption{Divide Node $i$}
    \label{alg:divide}
    \begin{algorithmic}[1]
        \Require $\Qcal$, $i$, $\Dcal = \{\x_j,\x_j^+\}_{j=1}^M$
        \State $\mathbf{v} = \{1,-1\}^n$
        \For{$k = N+1$ to $k = N+1+ 2^n$}
        \State $\hat{r}_k = \frac{\hat{r}_i}{2}$
        \State $\hat{\x}_k = \hat{\x}_i + \hat{r}_k{\mathbf{v}(k-N)}$
        \State $\x_k = \argmin_{\x \in \Dcal}\norm{\hat{\x}_k - \x}_{\infty}$
        \State $r_k = \hat{r}_k + \norm{\hat{\x}_k - \x_k}_{\infty}$
        \State Add leaf node $\Q(\hat{r}_k,\hat{\x}_k, r_k, \{\x_k, \x_k^+\}, s_k)$ to node i
        \State $s_k = 1$ in $\Q(\hat{r}_k,\hat{\x}_k, r_k, \{\x_k, \x_k^+\}, s_k)$
        \EndFor 
    \end{algorithmic}
    \Return $\Qcal$
\end{algorithm}

For any node $i\in \mathbb{Z}_1^N$ in $\Qcal,$ the partition $B_{\hat{r}_i,\infty}(\xhat_i)$ is covered by a hypercube about a sample from $\Dcal,$ $B_{r_i,\infty}(\x_i)\supseteq B_{\hat{r}_i,\infty}(\xhat_i).$  Therefore, $\Suc(B_{\hat{r}_i,\infty}(\xhat_i)) \subseteq \Suc(B_{r_i,\infty}(\x_i)) \subseteq B_{r_i,\infty}^+(\x_i)$, meaning the behavior of the sample pair $\{\x_i,\x_i^+\}$ can be used over approximate the behavior of the partition, $B_{\hat{r}_i,\infty}(\xhat_i)$. This principle later  guides the iterative partitioning and labeling of $\Xcal$ in \Cref{alg:findSet}.

\subsection{Algorithm}\label{sec:algTotal}

This section describes \Cref{alg:findSet}, which finds a \ac{pi} set of a system given sampled data. \Cref{alg:findSet} first assigns the entire constraint set as part of a candidate \ac{pi} set, $\hat{\Scal}$, i.e. $\hat\Scal \leftarrow\Xcal.$ The algorithm then iteratively partitions $\Xcal$, where each partition is a leaf node of the tree, $\Qcal.$ Hypercubes centered about samples from $\Dcal$ are used to cover the partitions. The partitions are then labeled based on the behavior of the relevant sampled data. In this way, the algorithm iteratively prunes away partitions of $\hat{\Scal}$ which are not positively invariant or, if the behavior of a partition is not clear, divides the leaf node -- creating more partitions for finer examination.

The candidate \ac{pi} set, $\hat{\Scal},$ used is constructed from the leaf nodes of $\Qcal$ where $s=1.$

\begin{definition}\label{def:hatS} 
    Let $L_\Qcal$ denote the set of indices of the $N$ leaf nodes of $\Qcal$ with value $s = 1$. Define $\hat{\Scal} {=}  \cup_{k\in L_\Qcal} B_{\hat{r}_k,\infty}(\hat{\x}_k).$ 
\end{definition}

\Cref{alg:findSet} is initialized with $\Qcal$ where $s_i = 1$ for all nodes $i \in \mathbb{Z}_1^N$ -- assigning all partitions of $\Xcal$ (or an inner-approximation of $\Xcal$) as the initial candidate \ac{pi} set, $\hat\Scal_1.$  The Lipschitz constant of \eqref{eq:dynSys} with respect to the max norm ($L > 0$), the minimum hypercube partition radius ($\tau > 0$), and the data set $\Dcal = \{\x_j,\x_j^+\}_{j=1}^M$ also initialize \Cref{alg:findSet}. 

\begin{algorithm}
    \caption{Synthesize \ac{pi} Set}
    \label{alg:findSet}
    \begin{algorithmic}[1]
        \Require $\Qcal$, $\tau>0$, $L > 0$, $\Dcal= \{\x_j,\x_j^+\}_{j=1}^M$
        \State  $z = 1$, $\hat{\Scal}_z \leftarrow \cup_{k \in L_\Qcal B_{r_k,\infty}(\x_k)}$, $\hat{\Scal}_{z-1} \leftarrow \emptyset$
        \While{$\hat{\Scal}_{z} \neq \hat{\Scal}_{z-1}$}
        \State $\hat{\Scal}_{z+1} \leftarrow \cup_{k \in L_\Qcal} B_{\hat{r}_k,\infty}(\hat{\x}_k)$
        \For{$ \Q(\hat{r}_i, \hat{\x}_i, r_i, \{\x_i, \x_i^+\}, s_i=1) \in  \Qcal$}
        \State $B^+_{r_i,\infty}(\x_i) \defeq \{\y\vecdim{n}\mid\norm{\x_i^+-\y}_{\infty}\leq Lr_i\}$
        \If{$B^+_{r_i,\infty}(\x_i) \cap \hat{\Scal}_{z+1} = B^+_{r_i,\infty}(\x_i)$}
        \State $s_i=1$ in $ \Q(\hat{r}_i, \hat{\x}_i, r_i, \{\x_i, \x_i^+\}, s_i)$
        \ElsIf{$B^+_{r_i,\infty}(\x_i) \cap \hat{\Scal}_{z+1} = \emptyset$}
        \State  $s_i=0$ in $ \Q(\hat{r}_i, \hat{\x}_i, r_i, \{\x_i, \x_i^+\}, s_i)$
        \Else
        \If{$\frac{\hat{r}_i}{2} \geq \tau$}
        \State  $\Qcal$ = \Cref{alg:divide}($\Qcal, i, \Dcal$)
        \Else
        \State $s_i=-1$ in $ \Q(\hat{r}_i, \hat{\x}_i, r_i, \{\x_i, \x_i^+\}, s_i)$
        \EndIf
        \EndIf
        \State $\hat{\Scal}_{z+1} \leftarrow \cup_{k \in L_\Qcal} B_{\hat{r}_k,\infty}(\hat{\x}_k)$
        \EndFor
        \State $z =z +1$
        \EndWhile
        \State $\Scal \leftarrow \hat{\Scal}_z$
    \end{algorithmic}
    \Return $\Scal$
\end{algorithm}

At each iteration, \Cref{alg:findSet} examines the behavior of each leaf node in $\Qcal$ where $i\in L_\Qcal$. If $B^+_{r_i,\infty}(\x_i) \cap \hat{\Scal} = B^+_{r_i,\infty}(\x_i),$ then all of $\Suc(B_{r_i,\infty}(\x_i))\supseteq \Suc(B_{\hat{r}_i,\infty}(\xhat_i))$ remains within $\hat{\Scal}$, and the partition $B_{\hat{r}_i,\infty}(\xhat_i)$ remains within the candidate \ac{pi} set ($s_i {=} 1).$  If $B^+_{r_i,\infty}(\x_i) \cap \hat{\Scal} = \emptyset,$ then all of $\Suc(B_{r_i,\infty}(\x_i))\supseteq \Suc(B_{\hat{r}_i,\infty}(\xhat_i))$ leaves $\hat{\Scal}$, and the partition $B_{\hat{r}_i,\infty}(\xhat_i)$ is labeled as not a part of the candidate \ac{pi} set  $(s_i {=} 0).$ If $B^+_{r_i,\infty}(\x_i) \cap \hat{\Scal} \subset B^+_{r_i,\infty}(\x_i)$, then $\Suc(B_{r_i,\infty}(\x_i))\supseteq \Suc(B_{\hat{r}_i,\infty}(\xhat_i))$ maps to both the candidate \ac{pi} set and outside of it, so the partition behavior is unclear. The tree node is then either divided into $2^n$ leaf nodes (creating $2^n$ more partitions) via \Cref{alg:divide} for more information or, if the target radius of the divided partition is less than threshold $\tau$, the partition $B_{\hat{r}_i,\infty}(\xhat_i)$ is labeled as unknown $(s_i {=} -1).$ 

Note that a limitation of \Cref{alg:findSet} is scalability, as division of a node requires $2^n$ new nodes, where $n$ is the system dimension.

\Cref{alg:findSet} iteratively labels the partitions of $\Xcal$ that make up the candidate \ac{pi} set until the candidate \ac{pi} set is equal from one iteration to the next. At this point, the candidate \ac{pi} set is a true \ac{pi} set, as stated below. 

\begin{theorem}\label{thm:converge}
    Let \Cref{asmpt:Lipschitz} hold for \eqref{eq:dynSys} with respect to the max norm. Let $\Qcal$ be a tree data structure as defined by \Cref{def:Q}, where $\cup_{k\in L_\Qcal} B_{{r}_k,\infty}(\x_k) \subseteq \Xcal$, and let $\Qcal$, $L$, $\tau$, and $\Dcal$ be inputs to \Cref{alg:findSet}. \Cref{alg:findSet} will produce a \ac{pi} set, $\Scal = \cup_{k \in L_\Qcal} B_{r_k,\infty}(\x_k)$, in a finite number of steps.
\end{theorem}
This proof parallels \cite[Theorem 11]{strong2025certificates} with a slight adjustment made to respond to the behavior of the partitions being over-approximated by hypercubes about sampled data points. 

\begin{proof}
    \Cref{alg:findSet} terminates when no changes are made in $\hat{\Scal}_{j+1}$ after iterating over each node within it -- implying no node divisions or label changes. If $\hat{\Scal}_{j+1} {=} \emptyset$, it is \ac{pi} by definition. If $\hat{\Scal}_{j+1} {\neq} \emptyset$, then $\cup_{k\in L_\Qcal} B_{r_k,\infty}^+(\x_k) {\subseteq} \cup_{k\in L_\Qcal} B_{\hat{r}_k,\infty}(\hat{\x}_k)$. By \Cref{lem:overPart}, $B_{\hat{r}_k,\infty}(\hat{\x}_k) {\subseteq} B_{r_k,\infty}(\x_k)$ for all $k{\in} L_\Qcal$. Thus, $\Suc(B_{\hat{r}_k,\infty}(\hat{\x}_k)) \subseteq \Suc(B_{r_k,\infty}(\x_k)) \subseteq B^+_{r_k,\infty}(\x_k)$. A non-empty and non-changing $\hat{\Scal}_{j+1}$ implies $\cup_{k\in L_\Qcal} \Suc(B_{\hat{r}_k,\infty}(\hat{\x}_k)) {\subseteq} \cup_{k\in L_\Qcal} B_{r_k,\infty}^+(\x_k) {\subseteq} \cup_{k\in L_\Qcal} B_{\hat{r}_k,\infty}(\hat{\x}_k).$ By \Cref{lem:pointsInv},  $\cup_{k\in L_\Qcal} B_{\hat{r}_k,\infty}(\hat{\x}_k)$ is a \ac{pi} set.

    The minimum radius, $\tau$, for partitions of $\Xcal$ ensures that only a finite number of partitions can be formed across $\Xcal$. Thus, \Cref{alg:findSet} iterates a finite number of times. 
\end{proof}

\section{Sample Bounds}

This section aims to establish when \Cref{alg:findSet} will find a non-empty \ac{pi} set for two distinct sampling cases: deterministic sampling of $\{\x,\x^+\}$ pairs and randomly sampling $\{\x,\x^+\}$ pairs from a uniform distribution across the state constraint set.
Because \Cref{alg:findSet} uses \Cref{lem:pointsInv} to synthesize a \ac{pi} set, we must first establish a relationship between the dynamics of \eqref{eq:dynSys} and the \ac{pi} sets that can be found using \Cref{lem:pointsInv}.

\subsection{Set Coverage}\label{sec:characSet}

\Cref{lem:pointsInv} finds a \ac{pi} set that is a union of norm balls, $\cup_{i=1}^NB_{r_i}(\x_i)$. An alternate perspective is that $\cup_{i=1}^NB_{r_i}(\x_i)$ covers some unknown \ac{pi} set. The question is then what factors can we determine about the \ac{pi} set that $\cup_{i=1}^NB_{r_i}(\x_i)$ is capable of covering.

First, we bound the distance of any point $\z\in B_{r}^+(\x)$ with respect to the set $\Scal$, given that $B_{r}(\x)\subseteq\Scal$.
\begin{lemma}\label{lem:psiBound}
    Let $\Scal$ be a C-set that is $\lambda-$contractive under the dynamics of \eqref{eq:dynSys}. Let $\bar{u} = \max_{\norm{\bu}_p\leq 1} \psi_\Scal(\bu).$ If $B_r(\x) \subseteq \Scal$, then $\psi_\Scal(\z) \leq \lambda + L r \bar{u}$ for all $\z\in B_r^+(\x).$  
\end{lemma}
\begin{proof}
    For any $\z \in B_r^+(\x)$, $\z = \x^+ + Lr\bu$ where $\norm{\bu} \leq 1.$ By the properties of the Minkowski function, $\psi_\Scal(\z)=\psi_\Scal(\x^+ + Lr\bu) \leq \psi_\Scal(\x^+) + Lr\psi_\Scal(\bu).$ By the contractivity of $\Scal$ and the bound $\bar{u}$, this expression is bounded above by $\lambda + Lr\bar{u}.$
\end{proof}

Using \Cref{lem:psiBound}, a bound on the radius of the norm ball $B_r(\x)$ can be established to ensure that if $B_{r}(\x)\subseteq\Scal$, then $B_{r}^+(\x)\subseteq\rho\Scal,$ where $\lambda < \rho < 1.$ This fact can be leveraged to determine the radius of $r$ for which the data driven geometric conditions of positive invariance hold.

\begin{lemma}\label{lem:invSetRho}
    Let $\Scal$ be a C-set that is $\lambda-$contractive under the dynamics of~\eqref{eq:dynSys}. Let $\bar{u} = \max_{\norm{\bu}_p\leq 1} \psi_\Scal(\bu)$. Let $\{\x_i\}_{i=1}^N \in \rho\Scal$ such that $\rho\Scal \subseteq \cup_{i=1}^N B_{r_i}(\x_i)$. If $\lambda + L \bar{u} r_i \le \rho \leq 1-r_i\bar{u}$ holds for every $i \in \mathbb{Z}_1^N$, then  $\cup_{i=1}^N B_{r_i}(\x_i)\subseteq \Scal$ and $\cup_{i=1}^N B_{r_i}(\x_i)$ is a \ac{pi} set.
\end{lemma}
\begin{proof}
    To show $B_{r_i}(\x_i)\subseteq \Scal$, verify $\psi_\Scal(\z) \leq 1$ for all $\z\in B_{r_i}(\x_i)$. For any $\z \in B_{r_i}(\x_i),$ $\psi_\Scal(\z) = \psi_\Scal(\x_i + r_i\bu)$ where $\norm{\bu}\leq 1.$ 
    By the logic of \Cref{lem:psiBound}, this is bounded above by $\rho + r_i\bar{u}.$ By assumption $\rho + r_i\bar{u} \leq 1$, which implies $\psi_\Scal(\z) \leq 1$ for all $\z\in B_{r_i}(\x_i)$. Thus, $\cup_{i=1}^N B_{r_i}(\x_i)\subseteq \Scal.$
    
    

    By \Cref{lem:psiBound}, when $r_i \leq \frac{\rho - \lambda}{L \bar{u}}$, $\psi_\Scal(\z) \leq \rho$ for all $\z \in B_r^+(\x).$
    Therefore, $B_{r_i}^+(\x) \subseteq \rho\Scal$ for all $i\in\mathbb{Z}_1^n$. Consequently, $\cup_{i=1}^N B^+_{r_i}(\x_i)\subseteq \rho\Scal.$ Because $\rho\Scal \subseteq \cup_{i=1}^N B_{r_i}(\x_i)$ by assumption, $\cup_{i=1}^N B_{r_i}(\x_i)$ is a \ac{pi} set by \Cref{lem:pointsInv}.
\end{proof}

\subsection{Sample Bounds}

Results from \Cref{sec:characSet} are now leveraged to give a sample bound on \Cref{alg:findSet}. We consider two sampling cases: one where the samples are deterministically sampled across $\Xcal$ and one where samples were generated by drawing randomly from a uniform distribution across $\Xcal.$

When sampling is deterministic, \Cref{alg:findSet} reduces to a set coverage problem, where \Cref{alg:findSet} aims to cover some \ac{pi} set with the minimal number of hypercube partitions needed. This is reflected in the sampling bound below.

\begin{theorem}\label{thm:boundDeterministic}
    Consider \eqref{eq:dynSys} and let \Cref{asmpt:Lipschitz} hold with respect to the max norm. Let $\Scal\subseteq\Xcal$ be an unknown, nonempty C-set that is $\lambda-$contractive under the dynamics of \eqref{eq:dynSys} and define the set $\rho\Scal$, where $\lambda < \rho < 1$. Further, define $\bar{u} = \max_{\norm{\bu}_\infty\leq 1} \psi_\Scal(\bu)$. Let $\Qcal$ be a tree data structure as defined by \Cref{def:Q}, where $\rho\Scal \subseteq \cup_{k\in L_\Qcal} B_{\hat{r}_k,\infty}(\xhat_k) \subseteq \Xcal$. Let $\Qcal$, $L$, $\Dcal = \{\x_j,\x_j^+\}_{j=1}^M$, and $\tau$ be inputs to \Cref{alg:findSet}. Further, let the data set $\Dcal = \{\x_j,\x_j^+\}_{j=1}^M$ be deterministically sampled such that $\xhat = \x$ and $\hat{r} = r$ in \Cref{alg:divide}. If $\lambda + \tau L \bar{u} \leq \rho \leq 1 - \tau\bar{u}$ and 
    \begin{equation}\label{eq:coverDet}
        M \geq \tau^{-n} \text{vol}(\Xcal),
    \end{equation}
    \Cref{alg:findSet} will find the \ac{pi} set $\cup_{i\in L_\Qcal} B_{\hat{r}_i,\infty}(\hat{\x}_i)\supseteq\rho\Scal$.
\end{theorem}
\begin{proof}
    By \cite[Proposition 4.2.10, Remark 4.2.12]{vershynin2018high}, $M$ is greater than the lower bound of samples needed for an $\epsilon-$net of hypercubes with radius $\tau$ to cover $\Xcal$. If \Cref{alg:findSet} progresses so that each partition is of radius $\tau$, there will exist indices $L_\rho{\subseteq} \mathbb{Z}_1^M$ such that $\{\hat{\x}_i\}_{i\in L_\rho}\in\rho\Scal$ creates an $\epsilon$-net, $\cup_{i\in L_\rho}B_{\tau,\infty}(\hat{\x}_i) \supseteq \rho\Scal$. By \Cref{lem:invSetRho}, $\cup_{i\in L_\rho}B_{\tau,\infty}(\hat{\x}_i)$ is a \ac{pi} set.

    In \Cref{alg:findSet},  $\hat{\Scal}_1 = \cup_{k \in L_\Qcal} B_{\hat{r},\infty}(\xhat_k) \supseteq \rho\Scal$ by assumption. For any $z{>} 1$, $\hat{\Scal}_z{\supseteq} \rho\Scal$ because, for any $\hat{\x}_k\in\rho\Scal$, $B^+_{r_k,\infty}(\x_k) \cap \hat{\Scal}_{z+1} \neq \emptyset$ by $\lambda$-contractivity. 
    At termination, \Cref{alg:findSet} will determine a \ac{pi} set $\cup_{k\in L_\Qcal}B_{\tau,\infty}(\hat{\x}_k)\supseteq \cup_{i\in L_\rho}B_{\tau,\infty}(\hat{\x}_i) \supseteq\rho\Scal.$
\end{proof}

\Cref{thm:probNonEmpty} considers the case where $\Dcal$ is generated by states being randomly sampled from a uniform distribution across $\Xcal$. Based on the density of the samples and the characteristics of \eqref{eq:dynSys}, it gives probabilistic guarantees of \Cref{alg:findSet} finding a non-empty \ac{pi} set.

\begin{theorem}\label{thm:probNonEmpty}

    Consider \eqref{eq:dynSys} and let \Cref{asmpt:Lipschitz} hold with respect to the max norm. Let $\Scal\subseteq\Xcal$ be an unknown, non-empty C-set that is $\lambda-$contractive under the dynamics of \eqref{eq:dynSys} and define the set $\rho\Scal$, where $\lambda < \rho < 1$ and $\rho\Scal \oplus B_{\tau,\infty}(0)\subseteq \Scal$. Further, define $\bar{u} = \max_{\norm{\bu}_\infty\leq 1} \psi_\Scal(\bu)$. Let $\Qcal$ be a tree data structure as defined by \Cref{def:Q}, where $\rho\Scal \subseteq \cup_{k\in L_\Qcal} B_{{r}_k,\infty}(\x_k) \subseteq \Xcal$. Let $\Qcal$, $L$, $\Dcal = \{\x_j,\x_j^+\}_{j=1}^M$, and $\tau$ be inputs to \Cref{alg:findSet}, where $ \{\x_i\}_{i=1}^N$ in $\Dcal$ is sampled uniformly across, $\Xcal$.
    
    If $\lambda + 2\tau L \bar{u}   \le \rho \leq 1-2\tau\bar{u}$ and
    \begin{equation}\label{eq:sampleBoundAlg}
        M \geq \frac{\log(\frac{1}{\delta}) + \log(\text{vol}(\Xcal)) +n\log(\frac{1}{\tau})}{\log(\frac{1}{1 -\frac{\tau^n}{\text{vol}(\Xcal)}})},
    \end{equation}
    then \Cref{alg:findSet} will find a \ac{pi} set $\cup_{k\in L_\Qcal}B_{\hat{r}_k,\infty}(\xhat_k) \supseteq \rho\Scal$ with probability $1-\delta.$
\end{theorem}

\begin{proof}
    Define $\overline{\rho\Scal} {=} \rho\Scal {\oplus} B_{\tau,\infty}(0)$. By the logic of the proof of \cite[Lemma 3]{korda2020computing}, if \eqref{eq:sampleBoundAlg} holds, there is a $1{-}\delta$ probability $\{\x_i\}_{i=1}^M$ in $\Dcal$ produces an $\epsilon$-net of hypercubes with radius $2\tau$ covering $\Xcal$. 
    Since $\overline{\rho\Scal}{\subseteq} \Xcal$, there is at least a $1{-}\delta$ probability there exists an $\epsilon$-net $\{\x_i\}_{i\in L_{\overline{\rho}}}{\in}\overline{\rho\Scal}$.
    By \Cref{lem:invSetRho}, $\cup_{i\in L_{\overline{\rho}}}B_{2\tau,\infty}({\x}_i)\supseteq \overline{\rho\Scal}$ is a \ac{pi} set.

    Consider if \Cref{alg:findSet} progresses to each node having radius $\tau$. 
    The $\epsilon$-net of partitions covering $\rho\Scal$ is defined as $\{\hat{\x}_v\}_{v\in L_{\rho}}\in\rho\Scal$, where $\cup_{v\in L_{\rho}} B_{\tau,\infty}(\xhat_v) \supseteq \rho\Scal$ and any $B_{\tau,\infty}(\xhat_v) \subseteq \cup_{i\in L_{\overline{\rho}}}B_{2\tau,\infty}({\x}_i).$ 
    For any  $v\in L_\rho$, $\min_{i\in L_{\overline{\rho}}}\norm{\xhat_v - \x_i} \leq \tau$ and $r_v=\tau+\|\hat{\x}_v-{\x_i}\|_\infty\le 2\tau$. Thus, any $B_{\tau,\infty}(\hat{\x}_v)$ can be covered by some $B_{2\tau,\infty}(\x_i)$, and $\Suc(B_{\tau,\infty}(\hat{\x}_v)) \subseteq \Suc(B_{2\tau,\infty}(\x_i)) \subseteq B^+_{2\tau,\infty}(\x_i) \subseteq \rho\Scal \subseteq \cup_{v\in L_{\rho}} B_{\tau,\infty}(\hat{\x}_v).$ Therefore, $\cup_{i\in L_\rho}B_{\tau,\infty}(\hat{\x}_v)$ is a \ac{pi} set.

    By the same arguments as \Cref{thm:boundDeterministic}, $\cup_{k\in L_\Qcal}B_{\hat{r}_k,\infty}(\xhat_k) \supseteq \cup_{v\in L_\rho}B_{r_v,\infty}(\x_v) \supseteq \rho\Scal$ for each iteration of \Cref{alg:findSet}. Thus, there is a  $1-\delta$ probability that \Cref{alg:findSet} produces the \ac{pi} set $\cup_{k\in L_\Qcal}B_{\hat{r}_k,\infty}(\xhat_k)\supseteq \rho\Scal$.
\end{proof}

\Cref{thm:boundDeterministic} and \Cref{thm:probNonEmpty} consider the ``worst case" scenario in which \Cref{alg:findSet} must partition all of $\Xcal$ to the smallest possible radius, $\tau$. Because \Cref{alg:findSet} adaptively partitions the space based on available samples, it's entirely possible for it to produce a \ac{pi} set with much less samples.

\section{Numerical examples}

\Cref{alg:findSet} was tested on an unmodeled, two-dimensional linear and an unmodeled, two-dimensional nonlinear system. 

\subsection{Unmodeled linear system}

\begin{figure}[t]
    \centering
    \includegraphics[width=0.9\linewidth]{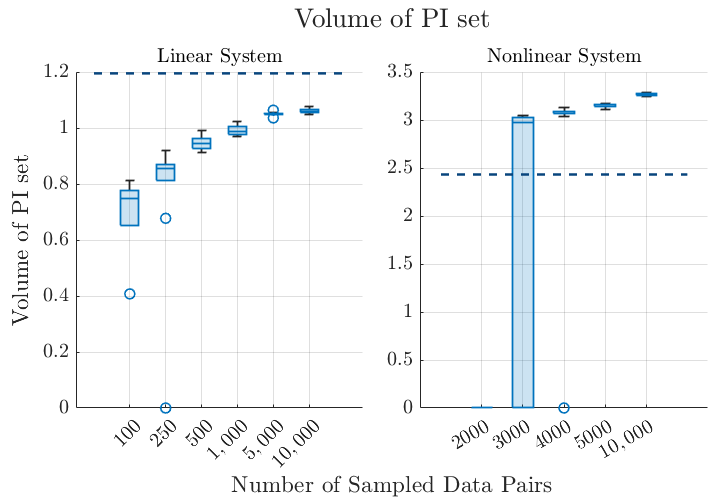}
    \caption{The volume of the \ac{pi} set found using \Cref{alg:findSet} is compared with the number samples used within the data set. The dotted black line shows the volume of the maximal \ac{pi} set found using model based methods.}
    \label{fig:boxSamples}
\end{figure}

Consider the linear system
\begin{flalign}\label{eq:exLinSys}
    \x^+ =\bmat{0.2200 & 0.4013 \\ -0.5364 & 0.2109}\x,
\end{flalign}
where $\Xcal: [-0.25,1]\times[-1, 0.25]$ and $L_{\infty} \geq 0.8225.$ Data sets were uniformly, randomly sampled across $\Xcal$ with varied sample sizes: $M = [100, 250, 500, 1000, 5000, 1e4].$ For each sample size, ten data sets were created. \Cref{alg:findSet} was applied to each data set with $\tau = 0.01$. \Cref{fig:boxSamples} (left) shows a box plot of the volume of the \ac{pi} sets resulting from application of \Cref{alg:findSet}. The dashed line is the volume of the maximal \ac{pi} set found using MPT3 \cite{MPT3}. As the density of the data set increased, the \ac{pi} set produced had greater volume more consistently. \Cref{alg:findSet} was typically able to find a non-empty \ac{pi} set even with low sample sizes.

\begin{figure}[t] 
    \centering
  \subfloat[\label{lineEx_100}]{%
       \includegraphics[width=0.49\linewidth]{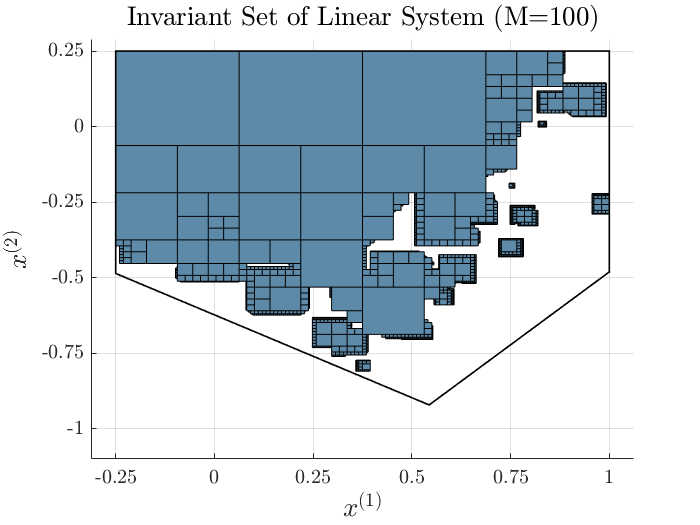}}
    \hfill
  \subfloat[\label{linEx_10000}]{%
        \includegraphics[width=0.49\linewidth]{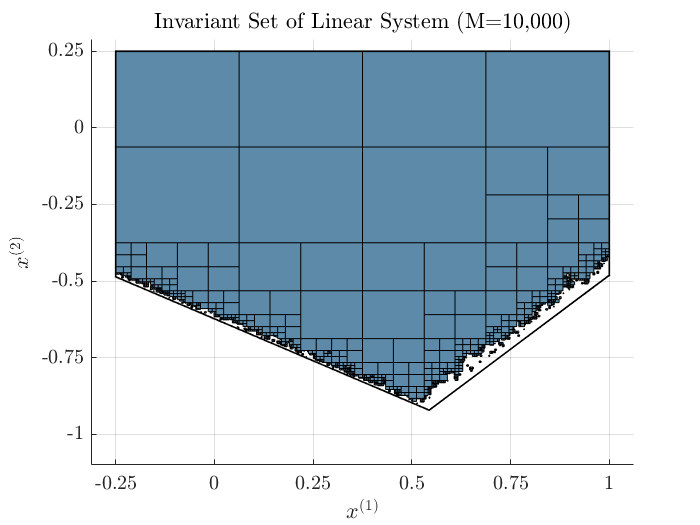}}
  \caption{The \ac{pi} set found by \Cref{alg:findSet} for \eqref{eq:exLinSys} is shown for data sets of (a) $M{=}100$ and (b) $M{=}10,000$ uniformly sampled points. Each blue square is a leaf node of the tree structure. The solid black line denotes the maximal \ac{pi} set of \eqref{eq:exLinSys} found by the model-based geometric method \cite{kerrigan2000robust} via \cite{MPT3}.}
  \label{fig:linEx_piSet} 
\end{figure}

 \Cref{fig:linEx_piSet} shows the maximal \ac{pi} set from MPT3 \cite{MPT3} as a solid black line; the resulting \ac{pi} set from application of \Cref{alg:findSet} ($\tau {=} 0.001$) for data sets with samples $M=100$ and $M=10,000$ are also shown in blue.  Each square represents a node, $i\in L_\Qcal$, of $\Qcal.$ The resulting \ac{pi} sets are not necessarily convex or even connected.
 
 \Cref{alg:findSet} can also be compared to the deterministic sampling scheme. For $M {=} 10,000$ ($\tau {=} 0.001$), \Cref{alg:findSet} produced a \ac{pi} set with volume $1.1554$. 
 In comparison, \cite[Alg 1]{strong2025certificates} ($\tau {=} 0.001$) found a \ac{pi} set with a volume of $1.1844$ using $M {=} 11,796$.


\subsection{Unmodeled nonlinear system}

Consider the nonlinear system
\begin{flalign}\label{eq:exNonlinSys}
    x_1^+ &= 0.5x_1 - 0.7x_2^2 \\
    x_2^+ &= 0.9x_2^3 + x_1x_2,
\end{flalign}
where $\Xcal = [-1,1]\times [-1, 1]$ and $L_{\infty} \geq 5.728.$ Data sets with $M = [2000, 3000, 4000, 5000, 1e4]$ samples were randomly sampled from a uniform distribution across $\Xcal.$ For each value of $M$, 10 different data sets were sampled. \Cref{alg:findSet} was applied to \eqref{eq:exNonlinSys} for each data set with $\tau = 0.01$. 

\begin{figure}[t] 
    \centering
  \subfloat[\label{nonlinEx_3000}]{%
       \includegraphics[width=0.49\linewidth]{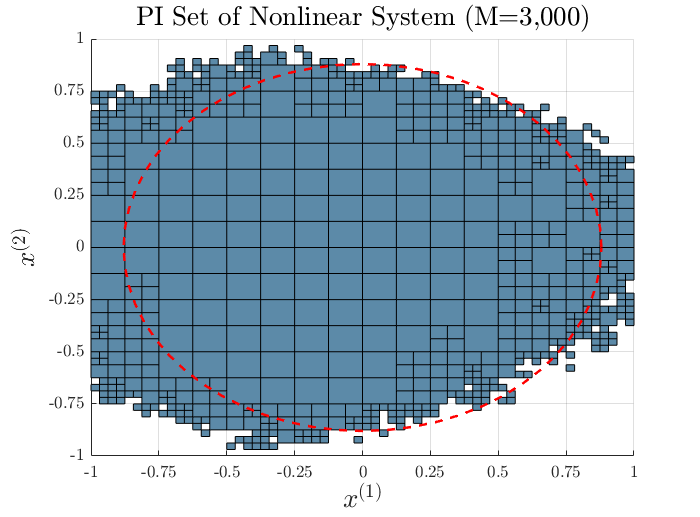}}
    \hfill
  \subfloat[\label{nonlinEx_10000}]{%
        \includegraphics[width=0.49\linewidth]{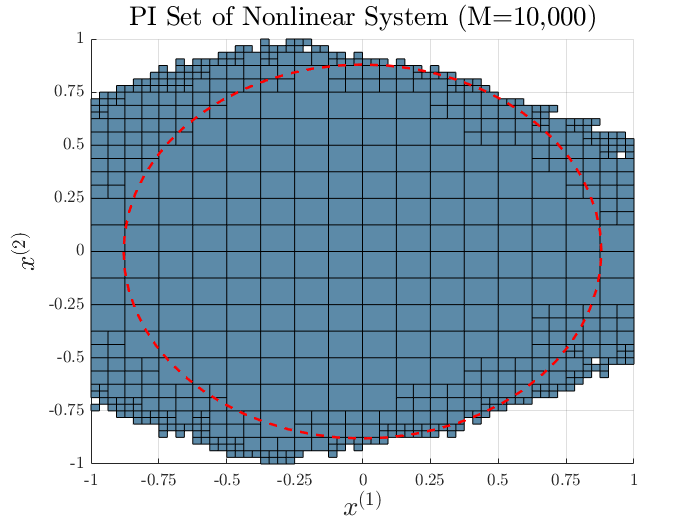}}
  \caption{The \ac{pi} set found by \Cref{alg:findSet} for \eqref{eq:exNonlinSys} is shown for data sets of (a) $M{=}3,000$ and (b) $M{=}10,000$ uniformly sampled points. Each blue square is a leaf node of the tree data structure. The red dashed curve denotes the \ac{pi} set induced via a barrier function determined using the system model.}
  \label{fig:nonlinEx_piSet} 
\end{figure}

\Cref{fig:boxSamples} (right) shows a box-plot of the resulting volume of the \ac{pi} set for these different data sets. As $M$ increases, the likelihood of \Cref{alg:findSet} producing a non-empty \ac{pi} set increases. Likewise, the volume of the \ac{pi} set increases. 
The largest \ac{pi} set produced had a volume of $3.286$ using $M=10,000.$ In comparison, the deterministic sampling scheme \cite[Alg. 1]{strong2025certificates} required $M = 2,178$ to produce a \ac{pi} set of volume $3.467.$ \Cref{fig:nonlinEx_piSet} shows an example of the \ac{pi} sets generated by \Cref{alg:findSet} for $M{=}3,000$ and $M {=} 10,000$ samples compared to sublevel set of the barrier function $V(\x) = x_1^2 + x_2^2.$ 


\section{Conclusions}

We created an algorithm that uses data driven geometric conditions of invariance to produce provable \ac{pi} sets for unmodeled systems. The only required information is the state constraint set, a data set of $\{\x,\x^+\}$ pairs sampled over the state constraint set, and a bound on the Lipschitz constant of the system. We also developed bounds on the number of samples needed for our algorithm to have a $1{-}\delta$ probability of producing a \ac{pi} set. Our numerical examples show the ability of our algorithm to produce  \ac{pi} sets that increase with volume as the number of samples increases.

\addtolength{\textheight}{-12cm}   






\bibliographystyle{IEEEtran}
\bibliography{IEEEabrv,biblio}

@string { ACC = "Amer. Control Conf."}

@string { SCL = "Systems \& Control Letters"}

@String{TAC 			   	= "IEEE Tran. Aut. Ctrl."}

@String{ACC				= "Amer. Ctrl. Conf."}

@String{CSL				= "IEEE Ctrl. Sys. Lett"}

@String{CDC59			= "59th IEEE Conf. Decis. Ctrl."}

@String{CDC62   		= "62nd IEEE Conf. Decis. Ctrl."}

@String{SCL				= "Sys. Ctrl. Lett."}

@String{SIAM_ctrl = "J. SIAM Control Optim."}

@String{PMLR = "Proc. Mach. Learn"}

@inbook{borrelli2017predictive,
  chapter={10},
  title={Predictive control for linear and hybrid systems},
  author={Borrelli, Francesco and Bemporad, Alberto and Morari, Manfred},
  year={2017},
  publisher={Cambridge University Press}
}

@article{alberto2007invariance,
  title={An invariance principle for nonlinear discrete autonomous dynamical systems},
  author={Alberto, Lus FC and Calliero, Tas R and Martins, Andre CP},
  journal=TAC,
  volume={52},
  number={4},
  pages={692--697},
  year={2007},
  publisher={IEEE}
}

@book{fitzpatrick2009advanced,
  title={Advanced calculus},
  author={Fitzpatrick, Patrick},
  volume={5},
  year={2009},
  publisher={Amer. Math. Soc}
}

@article{korda2020computing,
  title={Computing controlled invariant sets from data using convex optimization},
  author={Korda, Milan},
  journal=SIAM_ctrl,
  volume={58},
  number={5},
  pages={2871--2899},
  year={2020},
  publisher={SIAM}
}

@inbook{boyd2004convex,
  title={Convex optimization},
  chapter = {2},
  author={Boyd, Stephen P and Vandenberghe, Lieven},
  year={2004},
  publisher={Cambridge university press}
}

@article{dorea1999b,
  title={(A, B)-invariant polyhedral sets of linear discrete-time systems},
  author={Dorea, Carlos Eduardo Trabuco and Hennet, JC},
  journal={J. Optim. Theory Appl.},
  volume={103},
  pages={521--542},
  year={1999},
  publisher={Springer}
}

@book{lasalle2012stability,
  title={The stability and control of discrete processes},
  author={LaSalle, Joseph P},
  volume={62},
  year={2012},
  publisher={Springer Science \& Business Media}
}

@article{willems2005note,
  title={A note on persistency of excitation},
  author={Willems, Jan C and Rapisarda, Paolo and Markovsky, Ivan and De Moor, Bart LM},
  journal=SCL,
  volume={54},
  number={4},
  pages={325--329},
  year={2005},
  publisher={Elsevier}
}

@article{mulagaleti2021data,
  title={Data-driven synthesis of robust invariant sets and controllers},
  author={Mulagaleti, Sampath Kumar and Bemporad, Alberto and Zanon, Mario},
  journal=CSL,
  volume={6},
  pages={1676--1681},
  year={2021},
  publisher={IEEE}
}

@inproceedings{mejari2023direct,
  title={Direct data-driven computation of polytopic robust control invariant sets and state-feedback controllers},
  author={Mejari, Manas and Gupta, Ankit},
  booktitle=CDC62,
  pages={590--595},
  year={2023},
  organization={IEEE}
}

@article{wang2020scenario,
  title={Scenario-based set invariance verification for black-box nonlinear systems},
  author={Wang, Zheming and Jungers, Rapha{\"e}l M},
  journal=CSL,
  volume={5},
  number={1},
  pages={193--198},
  year={2020},
  publisher={IEEE}
}

@inproceedings{dellnitz2001algorithms,
  title={The algorithms behind GAIO—Set oriented numerical methods for dynamical systems},
  author={Dellnitz, Michael and Froyland, Gary and Junge, Oliver},
  booktitle={Ergodic theory, analysis, and efficient simulation of dynamical systems},
  pages={145--174},
  year={2001},
  organization={Springer}
}

@InProceedings {MPT3,
    author={M. Herceg and M. Kvasnica and C.N. Jones and M. Morari},
    title={{Multi-Parametric Toolbox 3.0}},
    booktitle={Proc.~of the European Control Conference},
    year={2013},
    address={Z\"urich, Switzerland},
    pages = {502--510}
}

@article{gilbert1991linear,
  title={Linear systems with state and control constraints: The theory and application of maximal output admissible sets},
  author={Gilbert, Elmer G and Tan, K Tin},
  journal=TAC,
  volume={36},
  number={9},
  pages={1008--1020},
  year={1991},
  publisher={IEEE}
}

@article{jin2023robust,
  title={Robust data-driven control barrier functions for unknown continuous control affine systems},
  author={Jin, Zeyuan and Khajenejad, Mohammad and Yong, Sze Zheng},
  journal=CSL,
  volume={7},
  pages={1309--1314},
  year={2023},
  publisher={IEEE}
}

@inproceedings{strong2025data,
  title={Data Driven Synthesis of Invariant Sets for Unmodeled Lipschitz Dynamical Systems using a Tree Data Structure},
  author={Strong, Amy K and Kashani, Ali and Danielson, Claus and Bridgeman, Leila J},
  booktitle=ACC,
  pages={2566--2571},
  year={2025},
  organization={IEEE}
}

@inproceedings{strong2025certificates,
  title={Data-driven certificates of constraint enforcement and stability for unmodeled, discrete dynamical systems using tree data structures},
  author={Strong, Amy K and Kashani, Ali and Danielson, Claus and Bridgeman, Leila J},
  booktitle={Symposium on Nonlinear Control Systems},
  year={2025},
  organization={IFAC}
}

@book{blanchini2015set,
  title={Set-theoretic methods in control},
  author={Blanchini, Franco and Miani, Stefano},
  year={2015},
  publisher={Springer}
}

@book{wainwright2019high,
  title={High-dimensional statistics: A non-asymptotic viewpoint},
  author={Wainwright, Martin J},
  volume={48},
  year={2019},
  publisher={Cambridge university press}
}

@book{kerrigan2000robust,
  title={Robust constraint satisfaction: Invariant sets and predictive control},
  author={Kerrigan, Eric C},
  year={2000},
  publisher={University of London}
}

@article{anand2023formally,
  title={Formally verified neural network control barrier certificates for unknown systems},
  author={Anand, Mahathi and Zamani, Majid},
  journal={IFAC-PapersOnLine},
  volume={56},
  number={2},
  pages={2431--2436},
  year={2023},
  publisher={Elsevier}
}

@inproceedings{jagtap2020control,
  title={Control barrier functions for unknown nonlinear systems using Gaussian processes},
  author={Jagtap, Pushpak and Pappas, George J and Zamani, Majid},
  booktitle=CDC59,
  pages={3699--3704},
  year={2020},
  organization={IEEE}
}

@book{vershynin2018high,
  title={High-dimensional probability: An introduction with applications in data science},
  author={Vershynin, Roman},
  volume={47},
  year={2018},
  publisher={Cambridge university press}
}

@inproceedings{edwards2024fossil,
  title={Fossil 2.0: Formal certificate synthesis for the verification and control of dynamical models},
  author={Edwards, Alec and Peruffo, Andrea and Abate, Alessandro},
  booktitle={Proceedings of the 27th ACM International Conference on Hybrid Systems: Computation and Control},
  pages={1--10},
  year={2024}
}

@article{zhang2024seev,
  title={Seev: Synthesis with efficient exact verification for relu neural barrier functions},
  author={Zhang, Hongchao and Qin, Zhizhen and Gao, Sicun and Clark, Andrew},
  journal={Advances in Neural Information Processing Systems},
  volume={37},
  pages={101367--101392},
  year={2024}
}

@inproceedings{richards2018lyapunov,
  title={The lyapunov neural network: Adaptive stability certification for safe learning of dynamical systems},
  author={Richards, Spencer M and Berkenkamp, Felix and Krause, Andreas},
  booktitle={Conference on robot learning},
  pages={466--476},
  year={2018},
  organization={PMLR}
}

@inproceedings{zhang2023efficient,
  title={Efficient sum of squares-based verification and construction of control barrier functions by sampling on algebraic varieties},
  author={Zhang, Hongchao and Li, Zhouchi and Dai, Hongkai and Clark, Andrew},
  booktitle=CDC62,
  pages={5384--5391},
  year={2023},
  organization={IEEE}
}

@inproceedings{robey2020learning,
  title={Learning control barrier functions from expert demonstrations},
  author={Robey, Alexander and Hu, Haimin and Lindemann, Lars and Zhang, Hanwen and Dimarogonas, Dimos V and Tu, Stephen and Matni, Nikolai},
  booktitle=CDC59,
  pages={3717--3724},
  year={2020},
  organization={IEEE}
}
\end{document}